\documentclass[preprint,
amsmath,amssymb, 
 aps, prd 
nofootinbib,
nobibnotes,
]{revtex4-1}

\usepackage{graphicx}
\usepackage{dcolumn}
\usepackage{bm}
\usepackage{soul}
\usepackage{placeins}
\usepackage{physics}
\usepackage[normalem]{ulem}
\usepackage{mathtools,slashed}
\usepackage{hyperref}
\usepackage{xcolor}
\usepackage[table]{xcolor}

\begin{document}


\title{Deep Inelastic Scattering and Proton Structure Functions in AdS/QCD models}

\author{Rafael A. Costa-Silva}
\author{Henrique Boschi-Filho}%
\email{rafaelcosta@pos.if.ufrj.br}
\email{rc-fis@outlook.com}
 \email{boschi@if.ufrj.br}
 \email{hboschi@gmail.com}
\affiliation{Instituto de Física, Universidade Federal do Rio de Janeiro, 21.941-909, Rio de Janeiro, RJ, Brazil
}%


\begin{abstract}
    In this work we study the holographical fermionic deep inelastic scattering in a deformed AdS model. Imposing that the  potentials that emerge from the Schrödinger-like equations for the  fermionic chiral left and right modes are SUSY partners to preserve  isospectrality, we implement a systematical truncation of such potentials. Then, we show that in the quadratic approximation of the deformed AdS model, the fermionic DIS and the proton structure functions $F_{1,2}(q^2,x)$ are identical with the results obtained from the soft wall model. Further, we obtain and compare high-order approximations for the fermionic potentials with the full solution of the deformed AdS model and show, as expected, that as greater the order of approximation, the better are the results for $F_{1,2}(q^2,x)$ when compared with experimental data explicitly for the Bjorken parameter values $x=0.85, 0.75$ and 0.65. 
\end{abstract}

\maketitle



\section{Introduction}

     Quantum chromodynamics (QCD), a non-Abelian Yang-Mills theory with $SU(3)$-color gauge group is the standard model description of strong interactions (for recent updates on QCD, see, {\sl e.g.}, \cite{Gross:2022hyw, Achenbach:2023pba}). Remarkably, it presents a negative beta function, and for this reason, at high-energies, the dimensionless running coupling constant shows up values smaller than one, and perturbative calculations can be safely done. On the other hand, for low energy processes, it becomes non-perturbative and other resources and techniques need be implemented to study its properties.

    A modern alternative approach to strong coupled systems, as low-energy QCD, is based on the AdS/CFT correspondence \cite{Maldacena:1997re, Gubser:1998bc, Witten:1998qj}, or more generally, on gauge/gravity dualities (for reviews, see, for instance, \cite{Aharony:1999ti, Petersen:1999zh, Nastase:2007kj, Nastase:2015wjb, Ammon:2015wua}). In its most well-known form, this correspondence relates a string theory in $AdS_5\times S^{5}$ to a conformal four dimensional gauge field theory with no gravity in Minkowski four dimensional spacetime. Since QCD is not a conformal theory, and if one wants to describe it using this correspondence, one needs to break its conformal invariance. In particular, a holographic approach to deep inelastic scattering (DIS) was proposed  in the seminal work of Polchinski and Strassler \cite{Polchinski:2002jw}. This process is a collision between a hadron (usually a proton) and a lepton, which allows the computation of the baryonic structure functions $F_1(x,q^2)$ and $F_2(x,q^2)$, where $x$ is the Bjorken variable representing the transferred momentum fraction in the collision and $q^2$ is the square of the momentum of the exchanged virtual photon   (for a review, see, {\sl e.g.}, \cite{Manohar:1992tz}). 
    
    In the last two decades, some models inspired on QCD phenomenology and the AdS/CFT correspondence emerged in order to describe strong interactions, and they are generically called AdS/QCD models. In particular, we may cite the soft wall model, \cite{Karch:2006pv, Kim:2007rt, Colangelo:2007pt, Grigoryan:2007my, Vega:2008af, Colangelo:2008us, Abidin:2009hr, Herzog:2006ra, Csaki:2006ji, Nakano:2006js,  Kim:2006uta, BallonBayona:2007vp,  Kwee:2007dd, He:2007juu, BallonBayona:2007qr,  Forkel:2007ru,  Batell:2008zm, Dubynskiy:2008di, dePaula:2008fp, Dietrich:2008up,  Forkel:2008un, Gursoy:2009zza,   Miranda:2009uw, Fujita:2009ca,  Afonin:2010fr, Branz:2010ub, Karch:2010eg, Gutsche:2011vb, Braga:2011wa, 
    He:2013qq, Li:2013oda,   Dudal:2015wfn, Rodrigues:2016kez, Mamo:2019mka}, the light-front holographic QCD \cite{Brodsky:2006uqa, Brodsky:2014yha,  deTeramond:2014asa} and the deformed AdS model \cite{Andreev:2006vy, Andreev:2006ct, Andreev:2006eh, Andreev:2006nw, Andreev:2007vn, Kajantie:2006hv, White:2007tu,  Csaki:2008dt, Giannuzzi:2009gh, Andreev:2009zk, Colangelo:2010pe, Li:2011hp, Yang:2015aia, Bernardini:2016hvx, Arefeva:2018hyo, Bruni:2018dqm, Giannuzzi:2019esi,  Bohra:2019ebj,  FolcoCapossoli:2019imm, Contreras:2021epz, MartinContreras:2021yfz, Nascimento:2023dzx, Nascimento:2025smf, Issifu:2026nay, Wang:2026dcy, Ovchinnikov:2026mve, Costa-Silva:2026eqp}. Previous holographic studies of DIS include {\sl e.g.} \cite{Polchinski:2002jw, BallonBayona:2007qr, Braga:2011wa, Deur:2018roz, Ahmady:2016ujw, Costa:2013uia, Watanabe:2012uc, Stoffers:2012zw, Koile:2014vca, Nishio:2011xz, FolcoCapossoli:2015hub, Mamo:2021cle, BallonBayona:2008zi, Chen:2016dlk, Watanabe:2019zny, BallonBayona:2010ae,
    Koile:2011aa} and in particular within the deformed AdS model in \cite{FolcoCapossoli:2020pks}.

    Here in this work, we will describe fermionic DIS,  within the deformed AdS model for large values of $x$, in the supergravity regime \cite{Polchinski:2002jw}. The fermionic field modes are not analytical in this model and we develop, following \cite{Costa-Silva:2026eqp}, systematical approximations for the fermionic potentials in order to obtain analytical solutions for this model, respecting the isospectrality of the chiral left and right modes, using supersymmetric quantum mechanics \cite{Witten:1981nf, Cooper:1994eh}. These analytical solutions for the fermionic fields in the deformed AdS model within the quadratic approximation can be mapped exactly in the dressed soft wall solution \cite{Braga:2011wa} with a convenient choice of the parameters of both models. Furthermore, we also compute numerical solutions for high-order approximations of the fermionic potentials and the full solutions to obtain the structure functions $F_1(x,q^2)$ and $F_2(x,q^2)$ of the proton in all of these cases, and compare these results with available experimental data.

    This work is organized as follows: in Sec. \ref{sec2} we briefly review the main properties of fermionic DIS. In Sec. \ref{sec3} we present the deformed AdS model, which describes the holographic interaction between a fermion and a vector field within our geometry. Furthermore, we write the equations of motion for the vector and fermionic fields and describe holographically the deep inelastic scattering, obtaining the proton structure functions $F_1(q^2,x)$ and $F_2(q^2,x)$. In Sec. \ref{sec4} we perform systematical truncations in the effective potentials of the fermionic fields, in order to preserve the isospectrality of the chiral left and right modes. We show that an approximation called quadratic approximation presents analytical solutions for the fermionic fields and also for the proton structure functions, we compare our results with the fermionic DIS in the dressed soft wall model. In Sec. \ref{sec5} we present our numerical analysis, presenting the proton structure functions for different values of $x$ and for different orders of approximations for the effective fermionic potentials. In Sec. \ref{sec6} we present our discussions and conclusions.


\section{Brief review of deep inelastic scattering}\label{sec2}

    Deep inelastic scattering is a quantum mechanical collision between a hadron (usually a proton), with momentum $P$ and a lepton through the exchange of a virtual photon, with momentum $q$. It is a great technique to obtain information about the internal structure of a proton, which experiments in the end of 1960's led to the discovery of quarks \cite{Breidenbach:1969kd} (for a review, see {\sl e.g.}, \cite{Riordan:1992hr}). The essential theoretical information about this scattering can be written in terms of the hadronic tensor, defined as \cite{Manohar:1992tz}
    \begin{equation}\label{hadronic1}
        W^{\mu\nu}=\frac{1}{4\pi}\sum_{s}\int{d^4ye^{iq\cdot y}}\bra{P,s}[J^{\mu}(y),J^{\nu}(0)]\ket{P,s},
    \end{equation}
    where $J^{\mu}$ is the electromagnetic hadronic current and $s$ is the spin of the proton. For an unpolarized target, taking into account Lorentz invariance, and writing only the symmetric terms that remain after contraction with the leptonic tensor, one can write
    \begin{equation}\label{hadronic2}
        W^{\mu\nu}=F_{1}(q^2,x)\bigg(\eta_{\mu\nu}-\frac{q^{\mu}q^{\nu}}{q^2}\bigg)+\frac{2x}{q}F_{2}(q^2,x)\bigg(P^{\mu}+\frac{q^{\mu}}{2x}\bigg)\bigg(P^{\nu}+\frac{q^{\nu}}{2x}\bigg).
    \end{equation}

    The functions $F_{1,2}(q^2,x)$ are the unpolarized DIS structure functions. The parameter $x=-q^2/2(P\cdot q)$ is the so-called Bjorken variable defined in the interval $0\leq x \leq1$. Notice that the lower $x$, the more inelastic the process is. On the other hand, $x\to1$ is the elastic limit. 
    
    If we insert a complete set of final hadronic states $\ket{X}$ with mass $M_X$ in equation (\ref{hadronic1}), one can rewrite $W^{\mu\nu}$ as
    \begin{equation}\label{hadronic3}
        W^{\mu\nu}=\frac{1}{4}\sum_{s}\sum_{X}\delta^{4}[M_{X}^{2}-(P+q)^2]\bra{P,s}J^{\mu}(0)\ket{X}\bra{X}J^{\nu}(0)\ket{P,s}.
    \end{equation}
    So, in order to obtain the proton structure functions $F_{1,2}(x,q)$, one can compute the matrix elements in this equation. We are going to evaluate these matrix elements from the holographic deformed AdS model below, following the prescriptions presented  in \cite{Polchinski:2002jw}. For a thorough review of non-holographic fermionic DIS see {\sl e.g.}, \cite{Manohar:1992tz}. 

\section{Deformed AdS model}\label{sec3}

    The deformed AdS/QCD model is described by an asymptotically AdS$_5$ space. With radius $R=1$, the metric can be written as 
    \begin{equation}\label{deformedmetric}
        ds^2=g_{mn}dx^{m}dx^{n}=e^{2A_{I}(z)}(dz^2+ \eta_{\mu\nu}dy^{\mu}dy^{\nu}),
    \end{equation}
    where the latin indices $m,n=0, 1, ..., 3, 5$, refers to the five-dimensional bulk space, separated in the holographic fifth coordinate $z$ and the Minkowski coordinates $x^\mu$ with greek indices $\mu,\nu=0, 1, 2, 3$, and metric $\eta_{\mu\nu}=\rm diag(-,+,+,+)$. Furthermore, the warp factor is defined by 
    \begin{equation}\label{warpfactor}
        A_{I}(z)=-\log(z)+\frac{k_{I}z^2}{2},
    \end{equation}
    with $k_{I}$ being two constants with dimension of mass squared, related to some QCD energy scales. Here, $I=(\gamma, f)$ refers to photon and fermions, respectively, so that each field couples differently with the static background as was discussed in the obtention of the pion form factor \cite{MartinContreras:2021yfz}. The reason for this is the fact that the warp factor signs need to be different for fermions and photons. If the signs are the same, or the photon solutions are not normalizable, or the fermions solutions presents spurious zero modes. In order to avoid these problems, we consider this model with  different signs for the warp factors for that different kinds of fields. In particular, we will work with the same absolute value for this constant, so we will write them from now on as
    \begin{equation}
        k_{f}=-k_{\gamma}=k.
    \end{equation}
    
    Notice that pure AdS$_5$ space is recovered in the limit where $k\rightarrow0$. DIS in this model was discussed previously in \cite{FolcoCapossoli:2020pks}. Here we present further aspects of this model and analysis with improvements which will also allow us to compare it with the  DIS in the dressed soft wall model \cite{Braga:2011wa}. This comparison will leave it clear the distinctions between these two models in general, and in particular in the calculations of the DIS structure functions. 

    In general, the holographic prescription to compute the matrix elements in equation (\ref{hadronic3}) at lowest order in the supergravity approximation is to identify them with the interaction action of the fields involved in the process \cite{Polchinski:2002jw, BallonBayona:2007qr, Braga:2011wa, 
    FolcoCapossoli:2020pks} 
    \begin{equation}\label{Jmu}
        \eta_{\mu}\bra{P+q,s_{X}|J^{\mu}(0)}\ket{P,s_i}=S_{int},
    \end{equation}
    where the interaction action of the photon and fermion fields is given by
    \begin{equation}\label{Sint}
        S_{int}=\int{dzd^4y\sqrt{-g_{f}}\mathcal{A^{\mu}}\bar{\Psi}_{X}\Gamma_{\mu}\Psi_i}.
    \end{equation}
    Here, $\mathcal{A}^{\mu}$ is the virtual photon field, $\Gamma_{\mu}$ is the Dirac gamma matrix in the curved space, $\Psi_{i/X}$ are, respectively, the initial and final baryonic states. So, in order to compute the matrix elements that appear in the hadronic tensor to obtain the proton structure functions, one needs to construct representations for these fields. Since we will compute this action on-shell, that implies in obtaining the normalizable solutions for the equations of motion of the involved fields in the deformed AdS geometry.

\subsection{Photon field}

 The action for a five dimensional massless gauge field $\mathcal{A}^{m}$, in a space which $g_{\gamma}$ is the determinant of the metric, is given by
    \begin{equation}
        S=-\int{d^{5}x\sqrt{-g_{\gamma}}\frac{1}{4}F^{mn}F_{mn}},
    \end{equation}
    where $F^{mn}=\partial^{m}\mathcal{A}^{n}-\partial^{n}\mathcal{A}^{m}$. The equations of motion obtained from this action are
    \begin{equation}
        \partial_{m}(\sqrt{-g_{\gamma}}F^{mn})=0.
    \end{equation}

    By using the gauge fixing
    \begin{equation}
        \partial_{\mu}\mathcal{A}^{\mu}+e^{-A_{\gamma}(z)}\partial_{z}(e^{A_{\gamma}(z)}\mathcal{A}_z)=0,
    \end{equation}
    with $A_\gamma (z)$ defined in equation (\ref{warpfactor}), one has
    \begin{equation}\label{eqgauge}
    \Box\mathcal{A}_{\mu}+A_{\gamma}'(z)\partial_{z}\mathcal{A}_{\mu}+\partial_{z}^{2}\mathcal{A}_{\mu}=0,
    \end{equation}
    \begin{equation}
        \Box\mathcal{A}_{z}-\partial_{z}(\partial_{\mu}\mathcal{A}^{\mu})=0, 
    \end{equation}
where prime denotes derivative with respect to the holographic coordinate $z$. 

    Now, without loss of generality, we will consider a particular polarization with transverse photons such that $\eta_{\mu}q^{\mu}=0$. So, as discussed in \cite{Polchinski:2002jw,BallonBayona:2007qr,Braga:2011wa}, only the component $\mathcal{A}_{\mu}$ will contribute in the scattering process that concerns us here. Then, one can write
    \begin{equation}
        \mathcal{A}_{\mu}=\eta_{\mu}e^{iq\cdot y}f(z,q).
    \end{equation}
Here, $q$ is the physical quadrimomentum of the photon. Since we will be interested in virtual photons, with the metric that we are working, $q^2>0$. The equation of motion for the holographic part is
    \begin{equation}\label{equationf}
        zf''(z)+(-1-kz^2)f'(z)-q^2zf(z)=0.
    \end{equation}
Defining $t=kz^2/2$ and $u(t)=f(z)$, $f'(z)=kzu'(t)$ and $f''(z)=ku'(t)+k^2z^2u''(t)$, and by making these substitutions in the above equation, one obtains 
    \begin{equation}
        tu''(t)+(0-t)u'(t)-\frac{q^2}{2k}u(t)=0, 
    \end{equation}
which is the confluent hypergeometric equation \cite{abramowitz1972handbook}. The linearly independent solutions for these parameters leads us to
    \begin{equation}
        \mathcal{A}_{\mu}(z,y)=C^{1}_{\mu}(y)\, G_{1, 2}^{2, 0} \left( -\frac{kz^2}{2} \,\middle|\, \begin{matrix} 1 - \frac{k q^2}{2} \\ 0, 1 \end{matrix} \right) +C^{2}_{\mu}\, kz^2~_{1}F_{1}\bigg(1+\frac{q^2}{2k},2,\frac{kz^2}{2}\bigg),
    \end{equation}
    where $G_{p,q}^{m,n} \left( z \,\middle|\, \begin{matrix} a_1 \cdots a_p \\ b_1 \cdots b_q \end{matrix} \right)$ is the Meijer G function and $_{1}F_{1}(a,b,c)$ is the Kummer confluent hypergeometric function. Imposing the boundary condition $\mathcal{A}_{\mu}(z,y)|_{z=0}=\eta_{\mu}e^{iq\cdot y}$, we have $C^{1}_{\mu}(y)=0$, and with the square integrable solution, one can write
\begin{equation}\label{fotonsoldeformed}
        \mathcal{A}_{\mu}(z,y)=\eta_{\mu}e^{iq\cdot y}\frac{1}{2}kz^2\, \Gamma\bigg(1+\frac{q^2}{2k}\bigg)\, \mathcal{U}\bigg(1+\frac{q^2}{2k},2,\frac{kz^2}{2}\bigg)\equiv\eta_{\mu}e^{iq\cdot y}B(z,q).
    \end{equation}
    Here, $\Gamma(x)$ is the Gamma function and $\mathcal{U}(a,b,c)$ is the Tricomi hypergeometric function. So, we now have the virtual photon solution to substitute in the interaction action.

\subsection{Fermion fields}

    The action for fermionic fields in deformed AdS background is
    \begin{equation}
        S=\int{d^4x dz\sqrt{-g_{f}}\, \bar{\Psi}(\slashed{D}-m_5)\Psi},
    \end{equation}
    where the covariant derivative is defined by \cite{FolcoCapossoli:2020pks}
       \begin{eqnarray}
\slashed{D}&=&g^{mn}e^{a}_{n}\gamma_{a}\bigg(\partial_{m}+\frac{1}{2}\omega^{bc}_{m}\Sigma_{bc}\bigg) \cr 
&=& e^{-A_f(z)}\gamma^{5}\partial_{5}+e^{-A_f(z)}\gamma^{\mu}\partial_{\mu}+2A_f'(z)\gamma^{5}.
    \end{eqnarray}

    By extremizing this action, one obtains
    \begin{equation}
        \big[e^{-A_{f}(z)}\gamma^{5}\partial_{5}+e^{-A_{f}(z)}\gamma^{\mu}\partial_{\mu}+2A_{f}'(z)\gamma^{5}-m_{5}\big]\Psi(x^{\mu},z)=0,
    \end{equation}
    where $\partial_{5}\equiv\partial_z$.
    It is useful to decompose the spinor in its chiral components,
     \begin{equation}
        \Psi(x^{\mu},z)=\bigg[\frac{1-\gamma^{5}}{2}f_{L}(z)+\frac{1+\gamma^{5}}{2}f_{R}(z)\bigg]\Psi_{(4)}(x^\mu), \label{Psi}
    \end{equation}
    where $\Psi_{(4)}(x^\mu)$ satisfies the usual Dirac equation $(\slashed{\partial}-M)\Psi_{(4)}(x^\mu)=0$ in flat four-dimensional spacetime. Notice that $M$ is the four-dimensional fermionic mass, and the chiral modes satisfy the relations
    \begin{equation}
        \gamma^{5}f_{L/R}=\mp f_{L/R},
        \label{relf1}
    \end{equation}
    and
    \begin{equation}
        \gamma^{\mu}\partial_{\mu}f_{R}=M f_{L}.
        \label{relf2}
    \end{equation}

    Witting the discrete modes for these chiral spinors, one can expand
    \begin{equation}
        \Psi_{L/R}(x^{\mu},z)=\sum_{n}f^{n}_{L/R}(x^{\mu})\, \chi^{n}_{L/R}(z).
    \end{equation}

    Performing a Bogoliubov transformation $\chi_{R/L}^{n}(z)=\psi^{n}(z)e^{-2A_{f}(z)}$, one can decouple the equations of motion for the chiral modes and obtain
    \begin{equation}\label{deformedpsieq}
        \bigg[-\frac{d^2}{dz^2}+V_{R/L}(z)\bigg]\psi^{n}_{R/L}(z)=M_{n}^{2}\psi_{R/L}^{2}(z),
    \end{equation}
    where $M_n$ are the four-dimensional discretized fermionic masses satisfying  $M_n^2=-P_n^2$ and the chral potentials are given by
    \begin{equation}\label{fullpotentials}
        V_{R/L}(z)=m_{5}^{2}e^{2A_{f}(z)}\pm m_{5}e^{A_{f}(z)}A_{f}'(z).
    \end{equation}

    The initial and final states of spinors can now be reconstructed from the solutions of equation (\ref{deformedpsieq}), respectively as
     \begin{equation}\label{psiideformed}
        \Psi_i=e^{iP\cdot y}e^{-kz^2}z^2\bigg[\bigg(\frac{1-\gamma^5}{2}\bigg)u_{s_i}(P)\psi_L^{i}(z)+\bigg(\frac{1+\gamma^5}{2}\bigg)u_{s_i}(P)\psi_R^{i}(z)\bigg],
    \end{equation}

     \begin{equation}\label{psixdeformed}
        \Psi_X=e^{iP_{X}\cdot y}e^{-kz^2}z^2\bigg[\bigg(\frac{1-\gamma^5}{2}\bigg)u_{s_X}(P_X)\psi_L^{X}(z)+\bigg(\frac{1+\gamma^5}{2}\bigg)u_{s_X}(P_X)\psi_R^{X}(z)\bigg].
        \end{equation}

    In general, the equations (\ref{deformedpsieq}) do not present analytical solutions for the potentials in (\ref{fullpotentials}), so in order to obtain the spectra for the fermions and write the solutions for fermionic modes, one needs to resort to numerical solutions. An exception is  the case of the quadratic approximation for these potentials presented in Sect. \ref{quadratic}, for which we find analytical solutions.


    \subsection{Fermionic DIS}

    With the solutions for the virtual photon field and the baryonic fields in hands, we are able to compute the interaction action \cite{FolcoCapossoli:2020pks},    
    \begin{align}\label{sint}
        S_{\rm{int}}&=g_{V}\int{dz d^{4}y\sqrt{-g_{f}}g^{\mu\nu}\mathcal{A}_{\mu}\bar{\Psi}_{X}e^{\alpha}_{\nu}\gamma_{\alpha}\Psi_{i}}\nonumber\\
        &=g_{V}\int{dzd^{4}y\sqrt{-g_{f}}e^{-2A_{f}(z)}\eta^{\mu\nu}\mathcal{A}_{\mu}\bar{\Psi}_{X}e^{A_{f}(z)}\delta^{\alpha}_{\mu}\gamma_{\alpha}\Psi_{i}}\nonumber\\
        &=g_{V}\int{dzd^{4}y\sqrt{-g_{f}}\eta^{\mu\nu}e^{-A_{f}(z)}\mathcal{A}_{\mu}\bar{\Psi}_{X}\gamma_{\mu}\Psi_{i}}\nonumber\\
        &=g_{V}\int{dzd^{4}y\frac{e^{2kz^2}}{z^4}\mathcal{A}^{\mu}\bar{\Psi}_{X}\gamma_{\mu}\Psi_{i}}.
    \end{align}

    Substituting the initial and final baryonic states from equations (\ref{psiideformed}) and (\ref{psixdeformed}) and the solution of the gauge field from equation (\ref{fotonsoldeformed}) in this action, we obtain
    \begin{equation}\label{Sintdelta}
        S_{\rm{int}}=g_{V}(2\pi)^4\delta^4(P_{X}-P-q)\eta^{\mu}(\bar{u}_{s_{X}}\gamma_{\mu}\hat{P}_{R}u_{s_{i}}\mathcal{I}_{L}+\bar{u}_{s_{X}}\gamma_{\mu}\hat{P}_{L}u_{s_{i}}\mathcal{I}_{R}),
    \end{equation}
    where the integrals $ \mathcal{I}_{R/L}$ are defined by 
    \begin{equation}\label{ILR}
        \mathcal{I}_{R/L}=\int{dz B(z,q)\, \psi^{X}_{R/L}(z,P_{X})\, \psi_{R/L}^{i}(z,P)}, 
    \end{equation}
while the holographic counterparts of the photon $B(z,q)$, and of the initial and final fermion fields  $\psi_{R/L}^{i}$, and $\psi^{X}_{R/L}$ are given by Eqs. 
\eqref{fotonsoldeformed}, \eqref{psiideformed}, and \eqref{psixdeformed}, respectively. 

    Then, using the expectation value of the four dimensional current \eqref{Jmu}, and the holographic interaction action \eqref{Sintdelta}, one gets
    \begin{equation}
        \eta_{\mu}\bra{P_{X}}J^{\mu}(q)\ket{P}=\frac{g_{\rm eff}}{2}\delta^{4}(P_{X}-P-q)\eta_{\mu}(\bar{u}_{s_{X}}\gamma_{\mu}\hat{P}_{R}u_{s_{i}}\mathcal{I}_{L}+\bar{u}_{s_{X}}\gamma_{\mu}\hat{P}_{L}u_{s_{i}}\mathcal{I}_{R})
    \end{equation}
    and
    \begin{equation}
        \eta_{\mu}\bra{P}J^{\mu}(q)\ket{P_{X}}=\frac{g_{\rm eff}}{2}\delta^{4}(P_{X}-P-q)\eta_{\mu}(\bar{u}_{s_{i}}\gamma_{\mu}\hat{P}_{R}u_{s_{X}}\mathcal{I}_{L}+\bar{u}_{s_{i}}\gamma_{\mu}\hat{P}_{L}u_{s_{X}}\mathcal{I}_{R}),
    \end{equation}
    where $g_{\rm eff}$ is an effective coupling related to $g_{V}$. Contracting the photon polarizations  $\eta_\mu$ with the hadronic tensor $W^{\mu\nu}$, Eq. \eqref{hadronic3}, and performing a summation over the spins, since we are interested in a spin independent scenario, one finds
    \begin{equation}
        \eta_{\mu}\eta_{\nu}W^{\mu\nu}=\frac{g_{\rm eff}^2}{4}\sum_{M_{X}^2}\delta(M_{X}^2-(P+q)^2)\bigg\{(\mathcal{I}_{L}^2+\mathcal{I}_{R}^2)\bigg[(P\cdot\eta)^2+\eta^2\bigg[-\frac{1}{2}(P^2+P\cdot q)+\mathcal{I}_L\mathcal{I}_R M_{X}^2M_0^2\bigg]\bigg]\bigg\}.
    \end{equation}

    In order to obtain the proton structure functions, one needs to perform the summation over the outgoing states. Following \cite{Polchinski:2002jw}, one has 
    \begin{equation}
        \delta(M_{X}^2-(P+q)^2)\propto \bigg(\frac{\partial M_{n}^2}{\partial n}\bigg)^{-1}.
    \end{equation}

    In our case, this delta function will account for $1/M_{X}^{2}$, with $M_{X}\equiv M_{X}(q^2,x)$ the mass of the effective final hadron related to the mass of the initial hadron
    \begin{equation}
        M_{X}^{2}(q^2,x)=M_{0}^{2}+q^2\bigg(\frac{1-x}{x}\bigg).
    \end{equation}

    Since we chose transversal photon polarization ($\eta\cdot q=0$), the hadronic tensor has the form
    \begin{equation}
        \eta_{\mu}\eta_{\nu}W^{\mu\nu}=\eta^{2}F_{1}(q^2,x)+\frac{2x}{q^2}(\eta\cdot P)F^{2}_{2}(q^2,x).
    \end{equation}

    Finally, we can read directly the spin independent proton structure functions
    \begin{align}\label{F1}
        &F_{1}(q^2,x)=\frac{g_{\rm eff}^2}{4}\bigg[M_{0}\sqrt{M_0^2+q^2\bigg(\frac{1-x}{x}\bigg)}\mathcal{I}_{L}\mathcal{I}_{R}+(\mathcal{I}_{L}^{2}+\mathcal{I}_{R}^{2})\bigg(\frac{q^2}{4x}+\frac{M_0^2}{2}\bigg)\bigg]\frac{1}{M_{X}^{2}},\\
        &F_{2}(q^2,x)=\frac{g_{\rm eff}^2}{8}\frac{q^2}{x}(\mathcal{I}_{L}^{2}+\mathcal{I}_{R}^{2})\frac{1}{M_{X}^{2}}\label{F2}, 
    \end{align}
in terms of the integrals (\ref{ILR}), which depend on the photon and fermion fields, \eqref{fotonsoldeformed}, \eqref{psiideformed}, and \eqref{psixdeformed}, respectively. To obtain the proton structure functions explicitly, we just need to compute these integrals. Since the chiral fermionic modes do not present analytical solutions, in general, the evaluation of these integrals are numerical, but for the quadratic approximation for the chiral potentials we find analytical solutions for these fields, as shown in next section.


\section{Truncations of the potentials for fermion fields}\label{sec4}

In this Section, we develop a systematic truncation of the chiral potentials, Eq. \eqref{fullpotentials}, keeping track of the isospectrality of the left and right modes. This is done with the help of supersymmetric quantum mechanics \cite{Witten:1981nf, Cooper:1994eh}, as discussed recently in \cite{Costa-Silva:2026eqp}. 

   We start defining the superpotential
    \begin{equation}
        W(z)=m_{5}e^{A_{f}(z)},
    \end{equation}
    with derivative
    \begin{equation}
        W'(z)=m_5e^{A_{f}(z)}A_{f}'(z).
    \end{equation}

    With these definitions, we see that the chiral potentials (\ref{fullpotentials}) can be written as
    \begin{equation}\label{relationsusy}
        V_{R/L}(z)=W^2(z)\pm W'(z).
    \end{equation}
This relation assures that the two chiral potentials be SUSY partners \cite{Witten:1981nf,Cooper:1994eh}. So, if one wants to obtain approximations of the chiral potentials, Eq. \eqref{fullpotentials}, one needs to perform truncations directly in the superpotential $W(z)$, because the relation (\ref{relationsusy}) we remain intact and the chiral modes, isospectral. Note that the superpotential, as defined for the deformed AdS model, only presents odd powers of $z$ \cite{Costa-Silva:2026eqp}
    \begin{equation}\label{W(z)}
        W(z)=\frac{m_5}{z}+\frac{m_5k}{2}z+\frac{m_5k^2}{8}z^3 + \frac{m_5k^3}{48}z^5 + \cdots
    \end{equation}

    So, since it appears as $W^2$ in the potentials, we only can obtain potentials with order $\mathcal{O}(z^{2j})$,  ($j=1,3,5,7,...$). After making some truncation in $W(z)$, we obtain approximations of the potentials that are SUSY partners, remaining the relation expected for the chiral modes. For example, by making the first order truncation in the superpotential, one obtains
    \begin{equation}
        W^{[1]}(z)=\frac{m_5}{z}+\frac{m_5k}{2}z.
    \end{equation}

    The corresponding chiral potentials are
    \begin{equation}\label{goodpotential1}
        V^{[1]}_{R/L}(z)=\frac{m_5(m_5\mp1)}{z^2}+km_{5}\bigg(m_5\pm\frac{1}{2}\bigg)+\frac{m_5^2k^2}{4}z^2.
    \end{equation}

    These potentials present up to order $z^2$ terms, and lead to analytical solutions. We will call this the quadratic approximation of the deformed AdS model and will develop this further below.


\subsection{Quadratic approximation and DIS}\label{quadratic}

    The quadratic approximation to the chiral potentials, Eq. \eqref{goodpotential1} substituted in the Schrödinger-like equation \eqref{deformedpsieq}, gives
        \begin{equation}\label{deformedpsieq2}
        \bigg[-\frac{d^2}{dz^2}+\frac{m_5(m_5\mp1)}{z^2}+km_{5}\bigg(m_5\pm\frac{1}{2}\bigg)+\frac{m_5^2k^2}{4}z^2\bigg]\psi^{n}_{R/L}(z)=M_{n}^{2}\psi_{R/L}^{2}(z).
    \end{equation}

   The solutions are (see the Appendix \ref{AppendA}): 
    \begin{equation}\label{psiR}
\psi_R^n(z)
=
\sqrt{
\frac{
\Gamma(n+1)
}{
\Gamma\left(n+m_5+\frac12\right)
}
}\sqrt{2\left(\dfrac{m_5 k}{2}\right)^{m_5+\frac12}}z^{m_5}
e^{-\frac{m_5 k}{4}z^2}L_n^{m_5-\frac12}\left(\frac{m_5 k}{2}z^2\right),
\end{equation}

\begin{equation}\label{psiL}
\psi_L^n(z)=
\sqrt{
\frac{\Gamma(n+1)
}{
\Gamma\left(n+m_5+\frac32\right)
}
}\sqrt{
2\left(\dfrac{m_5 k}{2}\right)^{m_5+\frac32}}z^{m_5+1}
e^{-\frac{m_5 k}{4}z^2}
L_n^{m_5+\frac12}
\left(\frac{m_5 k}{2}z^2\right).
\end{equation}

    As expected, these solutions are analytical. We can use these together with the one of the photon  (\ref{fotonsoldeformed}) to compute the DIS interaction action explicitly. The construction is the same as described in equation (\ref{sint}), and to obtain the proton structure functions in equations (\ref{F1}) and (\ref{F2}), we need just to compute the integrals (\ref{ILR}). To simplify our description, let us define the variable $\omega=kz^2/2$, so that
    \begin{equation}\label{integral}
        \mathcal{I}_{R/L}=C_{R/L}\int_{0}^{\infty}d\omega\bigg[ e^{-m_5\omega}\omega^{(m_{R/L}-1)}\,\mathcal{U}\bigg(1+\frac{q^2}{2k},2,\omega\bigg)L_{n}^{m_{R/L}-2}(m_5\omega)\bigg],
    \end{equation}
    where $m_{R/L}=m_5+2\mp 1/2$, and the constants $C_{R/L}$ are 
    \begin{equation}\label{constant}
        C_{R/L}=m_5^{(m_{R/L}-1)}\Gamma\bigg(1+\frac{q^2}{2k}\bigg)\sqrt{\frac{\Gamma(n+1)}{\Gamma(n+m_{R/L}-1)\Gamma(m_{R/L}-1)}}.
    \end{equation}

    At this point, one can see that the above calculation for the fermionic DIS in the deformed AdS model within the quadratic approximation can be related to the results found by in \cite{Braga:2011wa} for the dressed soft wall models. In particular, the  equations \eqref{integral} and \eqref{constant} reduce to Eqs. (3.20) and (3.21) of Ref. \cite{Braga:2011wa} when we put $m_5=1$ in our case and set $\kappa^2=k/2$, where $\kappa^2$ is the constant that appears in the dilaton of the dressed soft wall model. This relation occurs only for $m_5=1$ as our model is currently developed. But if we set $k_{\gamma}=-k$ and $k_{f}=k/m_5$, we obtain a full relation between the models for any choice of $m_5$.

    In order to compute the integrals in (\ref{integral}), we can expand the associated Laguerre polynomials $L_n^m(x)$ in series, to obtain
    \begin{equation}
        \mathcal{I}_{R/L}=C_{R/L}\sum_{j=0}^{n}\frac{(-m_5)^j\Gamma(n+m_{R/L}-1)}{j!(n-j)!\Gamma(j+m_{R/L}-1)}\int_{0}^{\infty}d\omega\bigg[\omega^{(m_{R/L}+j-1)}e^{-m_5\omega}\mathcal{U}\bigg(1+\frac{q^2}{2k},2,\omega\bigg)\bigg].
    \end{equation}

    These integrals present analytical solutions in terms of the Gauss hypergeometric function $_2F_1(\alpha,\beta;\gamma;z)$ \cite{Gradshteyn:1943cpj}, 
    \begin{equation}
        \int_{0}^{\infty}d\omega\big[\omega^{(b-1)}e^{-s\omega}\mathcal{U}(a,c,\omega)\big]=\frac{\Gamma(b)\Gamma(b-c+1)}{\Gamma(a+b-c+1)}\,_2F_1(b,b-c+1;a+b-c+1;1-s).
    \end{equation}
    In our case, $b=m_{R/L}+j$, $a=1+q^2/2k$, $c=2$ and $s=m_5$. So, our integral becomes a finite summation
    \begin{align}
        \mathcal{I}_{R/L}\; = & \;\; C_{R/L}\sum_{j=0}^{n}\frac{(-m_5)^{j}\Gamma(n+m_{R/L}-1)\Gamma(m_{R/L}+j)}{j!(n-j)!\Gamma\big(\frac{q^2}{2k}+m_{R/L}+j\big)}\,\nonumber\\
        &\times\;\;_2F_1 \bigg(m_{R/L}+j,m_{R/L}+j-1;\frac{q^2}{2k}+m_{R/L}+j;1-m_5\bigg).
    \end{align}
    This summation can be performed in order to obtain a closed analytical value to our integral $\mathcal{I}_{R/L}$. Performing this sum, we obtain
    \begin{align}\label{closedIRL}
        \mathcal{I}_{R/L}\, = & \;\;  m_5^{\,m_{R/L}-1}\,\frac{q^2}{2k}\,
\frac{\Gamma\!\left(\frac{q^2}{2k}+n\right)\Gamma\!\left(m_{R/L}\right)}
     {\Gamma\!\left(\frac{q^2}{2k}+m_{R/L}+n\right)}
\sqrt{\frac{\Gamma\!\left(n+m_{R/L}-1\right)}{n!\,\Gamma\!\left(m_{R/L}-1\right)}}\;\nonumber\\
    &\times\;\; _2F_1\!\left(m_{R/L},\,m_{R/L}+n-1;\,\frac{q^2}{2k}+m_{R/L}+n;\,1-m_5\right).
    \end{align}
 
After calculating the structure functions 
     $F_{1,2}$ in the quadratic approximation for the deformed AdS model, one might ask what happens to these functions when we consider higher order approximations to the chiral potentials of this model? In the next section, we examine this question with the numerical evaluation of $F_2$ for various approximations and compare them with available experimental data.


\section{Numerical results for the Structure functions of the Proton}\label{sec5}

    Now, we compute explicitly the proton structure functions for some approximations of the superpotential $W(z)$. We call the quadratic approximation $W^{[1]}(z)$, since it presents $V^{[1]}$ of order $z^2$. So, $W^{[3]}(z)$ and $W^{[5]}(z)$ will produce potentials $V^{[3]}(z)$ and $V^{[5]}(z)$, up to $\mathcal{O}(z^6)$ and $\mathcal{O}(z^{10})$, respectively. For this reason, these last approximations will be called sextic and decic approximations.

    Here, our main objective is to compare these approximations with the full potential, without approximations, as obtained in \cite{FolcoCapossoli:2020pks}. We expect that high-order approximations presents better results for the proton structure functions. In order to compare the solutions, we will plot the proton structure function $F_{2}(q^2,x)$ and compare the obtained results with experimental data \cite{WHITLOW1992475, BCDMS:1989qop}. 

\subsection{Quadratic and sextic approximations}

    The quadratic approximation, given by the chiral potentials in Eq. (\ref{goodpotential1}), present analytical solutions, discussed in the previous section. For high-order approximations, we need to use numerical resources. The next-order approximation is obtained from $W^{[3]}(z)$, given by
    \begin{equation}
        W^{[3]}(z)=\frac{m_5}{z}+\frac{m_5k}{2}z+\frac{m_5k^2}{8}z^3,
    \end{equation}
    which leads to the sextic approximation for the chiral potentials,
    \begin{equation}\label{sexticpotential}
         V_{R/L}^{[3]}(z)=\frac{m_5(m_5\mp1)}{z^2}+km_5\bigg(m_5\pm\frac{1}{2}\bigg) +k^2m_5\bigg(\frac{m_5}{2}\pm\frac{3}{8}\bigg)z^2 +\frac{m_5^2k^3}{8}z^4+\frac{m_5^2k^4}{64}z^6.
    \end{equation}

\begin{figure}[ht!]
\vskip0.5cm 
\centering
\includegraphics[width=16cm]{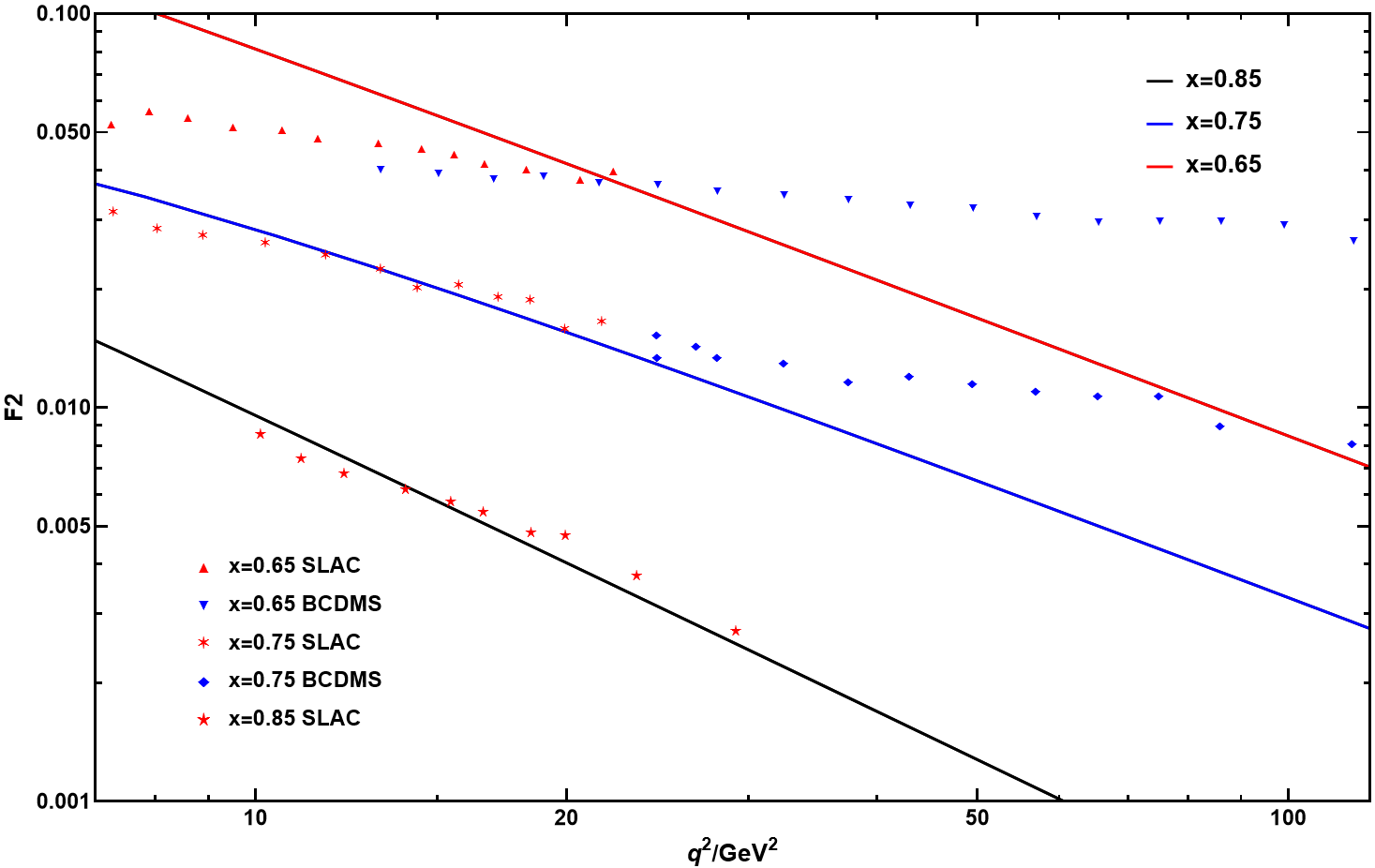}
\caption{Proton structure function $F_2(q^2,x)\times q^2$ obtained from the quadratic approximation of the deformed AdS model, equation (\ref{goodpotential1}). The lines represent our the results for $x=0.85$ (black),  $x=0.75$ (blue), and $x=0.65$ (red). The other values of the potential parameters and the effective coupling $g_{\rm eff}$ used in this plot are given in Table \ref{tab1}.}
\label{fig1}
\end{figure}

\begin{figure}[ht!]
\vskip0.5cm 
\centering
\includegraphics[width=16cm]{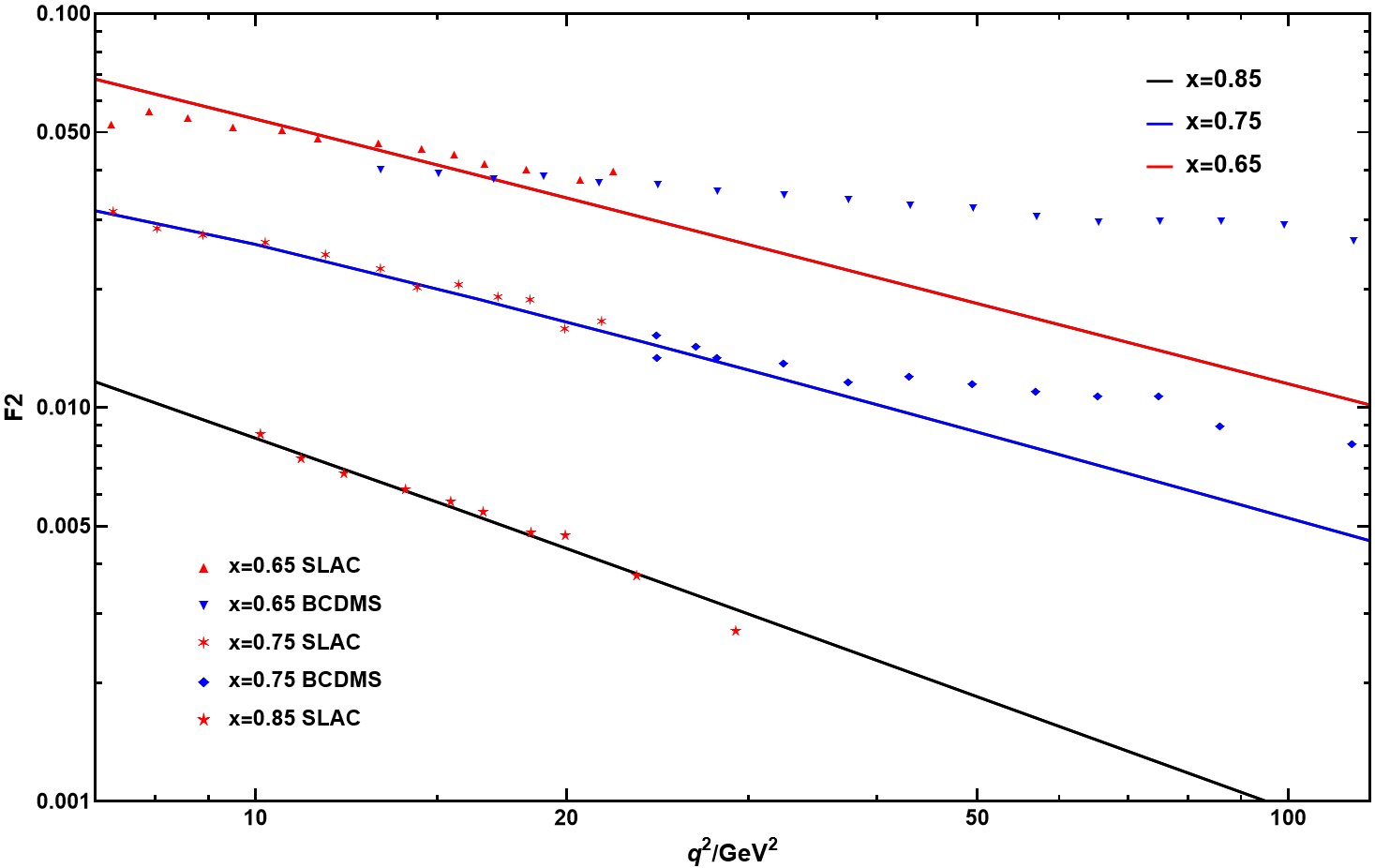}
\caption{Proton structure function $F_2(q^2,x)\times q^2$ obtained from the sextic approximation of the deformed AdS model, equation (\ref{sexticpotential}). The lines represent our the results for $x=0.85$ (black), $x=0.75$ (blue), and $x=0.65$ (red). The other values of the potential parameters and the effective coupling $g_{\rm eff}$ used in this plot are given in Table \ref{tab1}.}
\label{fig2}
\end{figure}

    In Table \ref{tab1}, we present the numerical parameters used in the quadratic and sextic approximations which reproduce the proton mass $M_{0}=938$ MeV as the hadronic ground state, and produce the curves for the proton structure functions, shown in Figs. \ref{fig1} and \ref{fig2}. We plot $F_2(q^2,x)$ for each case and compare its behavior for different values for the Bjorken parameter $x=0.85, 0.75, 0.65$,  present our results for the quadratic and sextic approximations, and comparisons with available experimental data \cite{WHITLOW1992475, BCDMS:1989qop}. 

\vskip 0.5cm
\begin{table}[h!]
\begin{tabular}{|c|c|c|c|c|}
\hline
\hline 

 Approximation & $x$ &$m_5$ &\, $k$ (GeV)$^2$ \, & \, $g_{\rm eff}^2$ 
\\ \hline

&\, 0.85 \,&\, 0.765 \, & \, 0.650$^2$ \, &   7.60\,\\
  Quadratic&0.75 &\, 0.510 \, & \, 0.924$^2$ \, & 8.80\,   \\
 &0.65 &\, 0.500 \, & \, 0.938$^2$ \, & 32.0\,   \\
 \hline
 &0.85 &\, 0.770 \, & \, 0.518$^2$ \, & 1.84\,   \\
  Sextic&0.75 &\, 0.565 \, & \, 0.636$^2$ \, & 2.80\,   \\
 &0.65 &\, 0.505 \, & \, 0.689$^2$ \, & 5.20\,   \\
 \hline\hline
\end{tabular}
\caption{Values of $m_5$, $k$ and $g_{\rm eff}^2$ for each value of the Bjorken parameter $x$ in the quadratic and sextic approximations of the deformed AdS model, equations (\ref{goodpotential1}) and (\ref{sexticpotential}), used in Figs. \ref{fig1} and \ref{fig2}.}
\label{tab1}
\end{table}   

\subsection{Decic approximation and full solution}

    The full solution is obtained by using the potentials in (\ref{fullpotentials}) and the decic approximation is obtained from $W^{[5]}(z)$
    \begin{equation}
        W^{[5]}(z)=\frac{m_5}{z}+\frac{m_5k}{2}z+\frac{m_5k^2}{8}z^3 + \frac{m_5k^3}{48}z^5,
    \end{equation}
    which leads us to the decic approximation of the chiral potentials
    \begin{equation}
\begin{aligned}\label{decicpotential}
V_{R/L}^{[5]}(z) ={}&
\frac{m_5(m_5\mp 1)}{z^2}
+ m_5 k\left(m_5\pm\frac{1}{2}\right)+m_5 k^2\left(\frac{m_5}{2}\pm\frac{3}{8}\right)z^2\\[4pt]&+m_5 k^3\left(\frac{m_5}{6}\pm\frac{5}{48}\right)z^4
+\frac{7m_5^2 k^4}{192}z^6
+\frac{m_5^2 k^5}{192}z^8
+\frac{m_5^2 k^6}{2304}z^{10}.
\end{aligned}
\end{equation}

\begin{figure}[h!]
\vskip0.5cm 
\centering
\includegraphics[width=16cm]{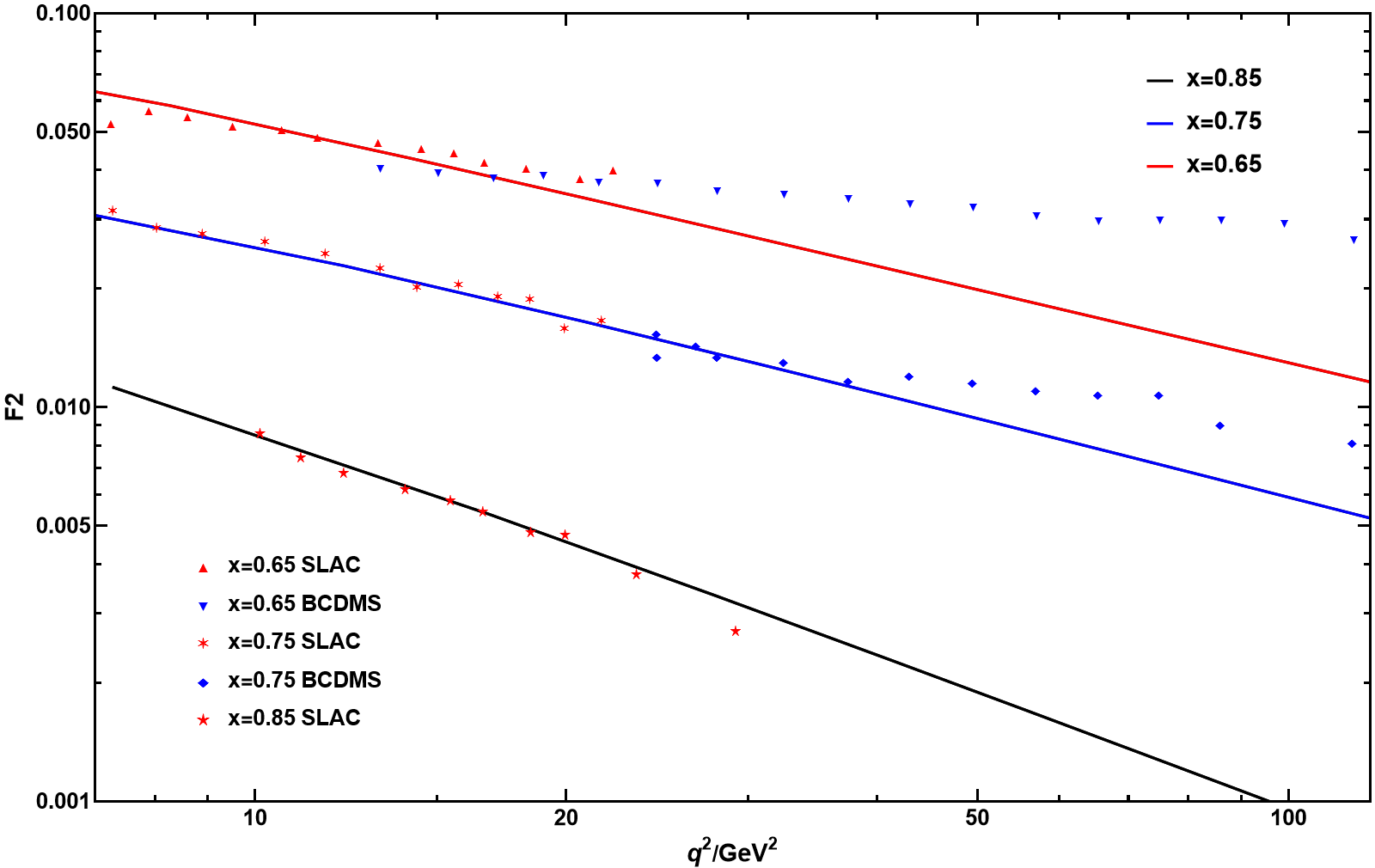}
\caption{Proton structure function $F_2(q^2,x)\times q^2$ obtained from the decic approximation of the deformed AdS model, equation (\ref{decicpotential}).  The lines represent our the results for $x=0.85$ (black), $x=0.75$ (blue), and $x=0.65$ (red). The other values of the potential parameters and the effective coupling $g_{\rm eff}$ used in this plot are given in Table \ref{tab2}.}
\label{fig3}
\end{figure}

\begin{figure}[h!]
\vskip0.5cm 
\centering
\includegraphics[width=16cm]{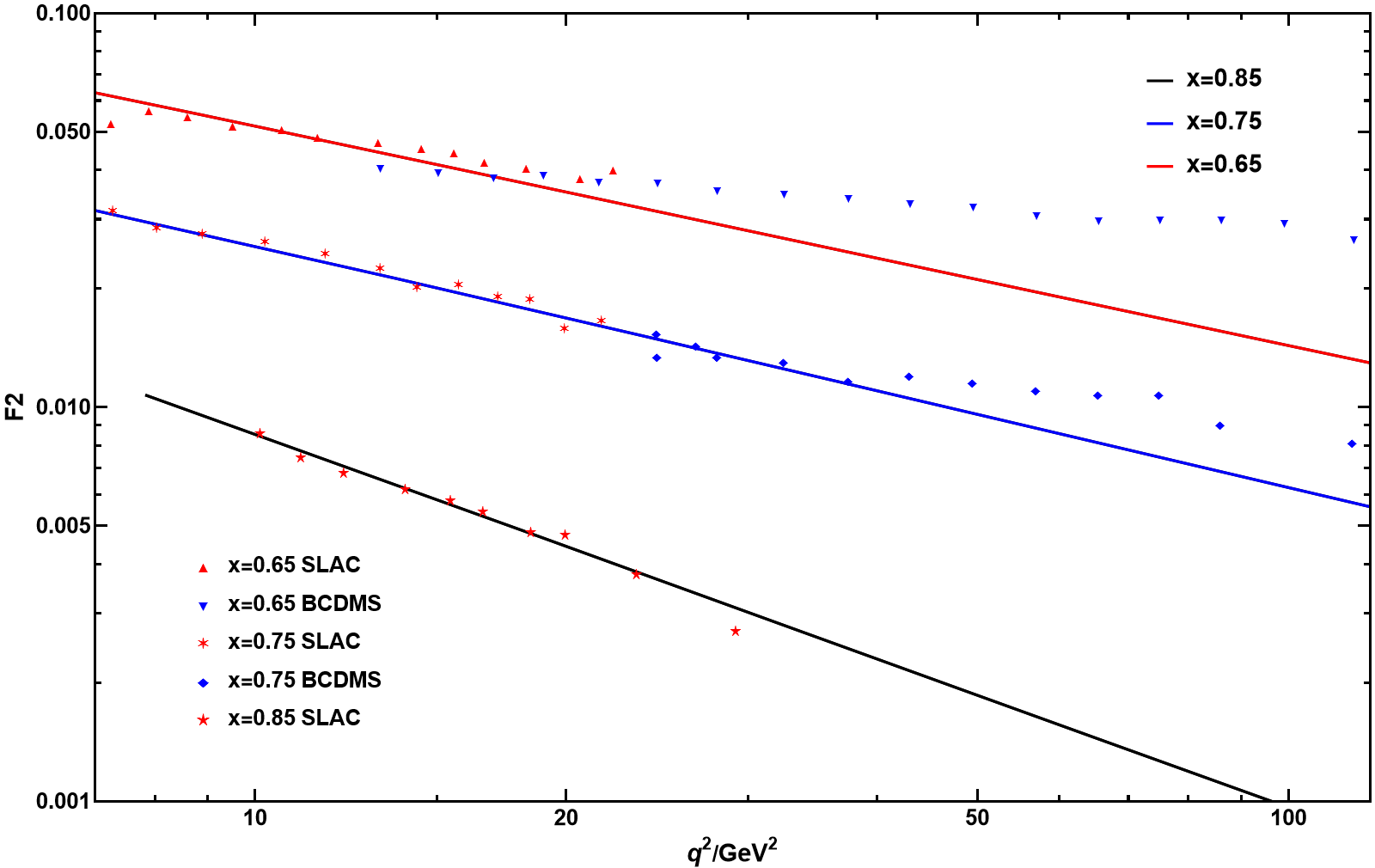}
\caption{Proton structure function $F_2(q^2,x)\times q^2$ obtained from the of the deformed AdS model with the full potentials, equation (\ref{fullpotentials}).  The lines represent our the results for $x=0.85$ (black), $x=0.75$ (blue), and $x=0.65$ (red). The other values of the potential parameters and the effective coupling $g_{\rm eff}$ used in this plot are given in Table \ref{tab2}.}
\label{fig4}
\end{figure}

    We present in Table \ref{tab2} the numerical values of the parameters used in the decic approximation and the full deformed AdS model which fit the proton mass $M_{0}=938$ MeV as the hadronic ground state, and obtain the proton structure functions. The results for the proton structure function $F_2(q^2,x)$ in the decic approximation and from the full solution of the deformed AdS, are presented for different values of the Bjorken parameter $x=0.85, 0.75, 0.65$ in Figures \ref{fig3} and \ref{fig4}, respectively, together with comparisons with experimental data \cite{WHITLOW1992475, BCDMS:1989qop}. 

\vskip 0.5cm
\begin{table}[h!]
\begin{tabular}{|c|c|c|c|c|}
\hline
\hline 

 Approximation & $x$ &$m_5$ &\, $k$ (GeV)$^2$ \, & \, $g_{\rm eff}^2$ 
\\ \hline

&\, 0.85 \,&\, 0.880 \, & \, 0.449$^2$ \, &   2.80\,\\
  Decic&0.75 &\, 0.565 \, & \, 0.600$^2$ \, & 2.08\,   \\
 &0.65 &\, 0.505 \, & \, 0.648$^2$ \, & 3.76\,   \\
 \hline
 &0.85 &\, 0.878 \, & 0.443$^2$\,  \, & 2.16\,   \\
  Full&0.75 &\, 0.565 \, & 0.583$^2$\,  \, & 1.68\,   \\
 &0.65 &\, 0.505 \, & 0.612$^2$\,  \, & 3.36\,   \\
 \hline\hline
\end{tabular}
\caption{Values of $m_5$, $k$ and $g_{\rm eff}^2$ for each value of the Bjorken parameter $x$ in the decic approximation of the deformed AdS model, equation (\ref{decicpotential}) and the full potential, equation (\ref{fullpotentials}), used in Figs. \ref{fig3} and \ref{fig4}.}
\label{tab2}
\end{table}

    The plots in Figures [\ref{fig1}-\ref{fig4}] show us that the greater the order of the approximation, the nearer our theoretical results for $F_{2}(q^2,x)$ become to the experimental data. The decic approximation looks also very similar to the results of the full deformed AdS model. Also, it is clear that our results for the structure function $F_2$ are better for the higher values of $x$. This can be understood since the supergravity approximation used here works better for larger values of $x$, as predicted in \cite{Polchinski:2002jw}.

\subsection{Comparison between approximations for chiral potentials with fixed $x$}

\begin{figure}[h!]
\vskip0.5cm 
\centering
\includegraphics[width=16cm]{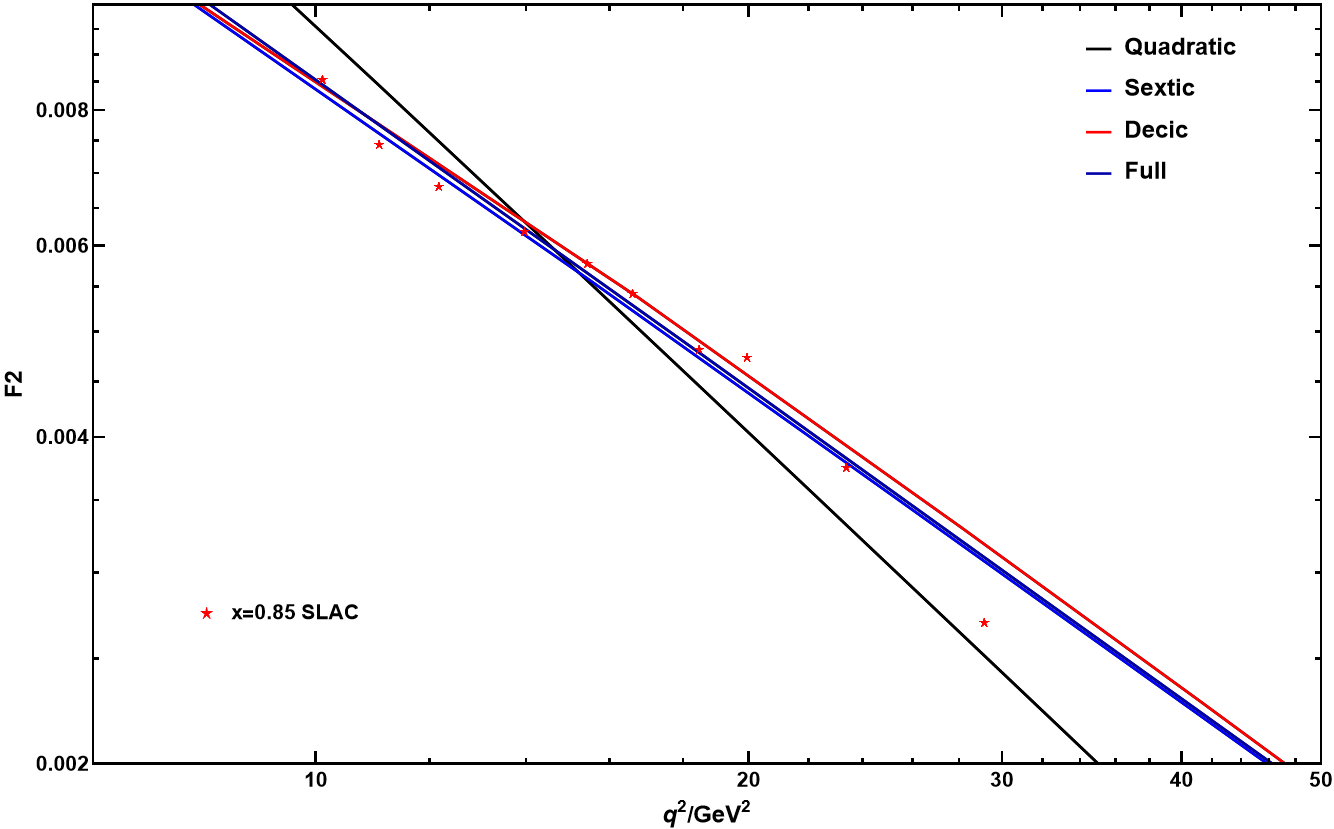}
\caption{Proton structure function $F_2(q^2,x)\times q^2$ obtained from the of the deformed AdS model for $x=0.85$. Black line - quadratic approximation (\ref{goodpotential1}), sextic approximation - blue line (\ref{sexticpotential}), for decic approximation - red line (\ref{decicpotential}), the full solution is given by the dark blue line (\ref{fullpotentials}).}
\label{fig5}
\end{figure}

\begin{figure}[h!]
\vskip0.5cm 
\centering
\includegraphics[width=16cm]{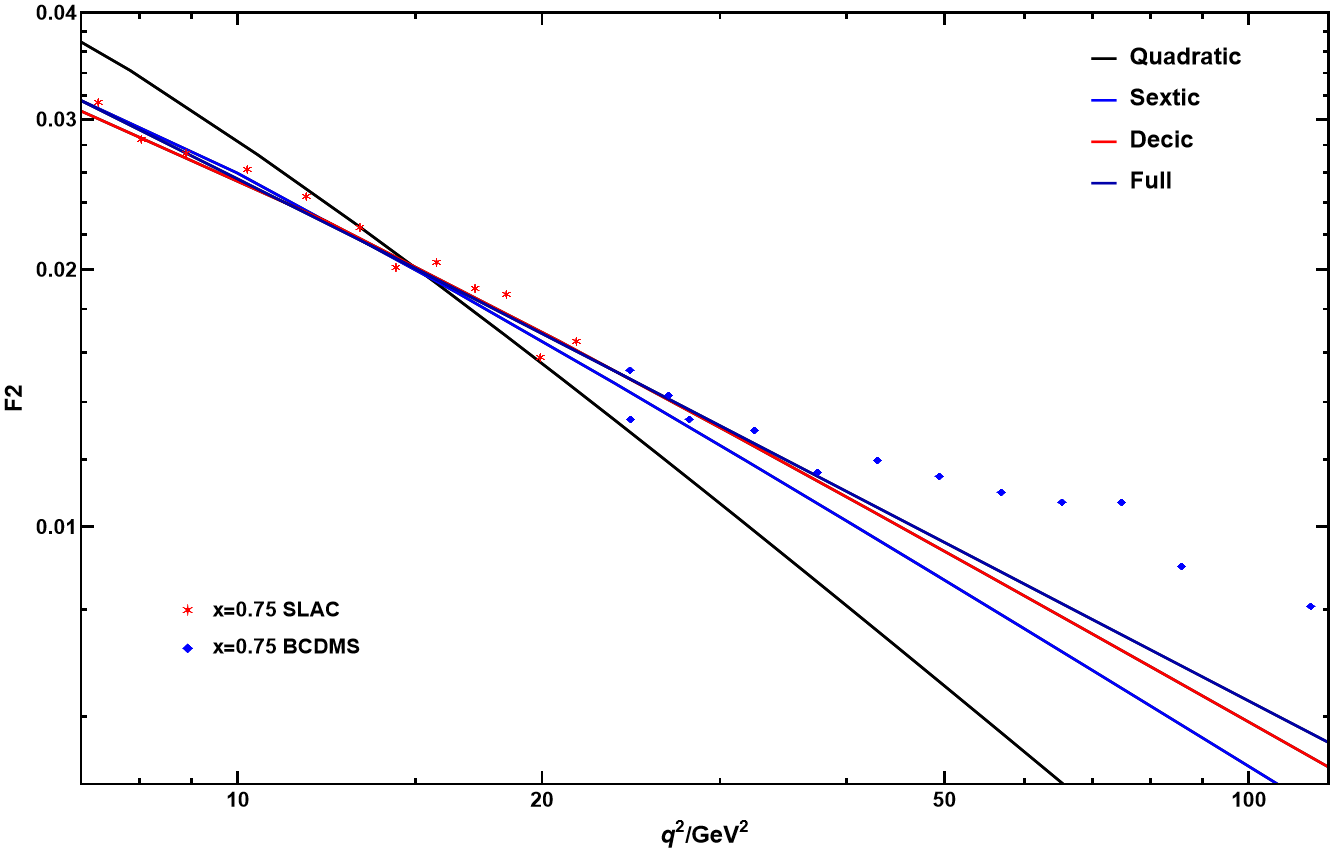}
\caption{Proton structure function $F_2(q^2,x)\times q^2$ obtained from the of the deformed AdS model for $x=0.75$. Black line - quadratic approximation (\ref{goodpotential1}), sextic approximation - blue line (\ref{sexticpotential}), for decic approximation - red line (\ref{decicpotential}), the full solution is given by the dark blue line (\ref{fullpotentials}).}
\label{fig6}
\end{figure}

\begin{figure}[h!]
\vskip0.5cm 
\centering
\includegraphics[width=16cm]{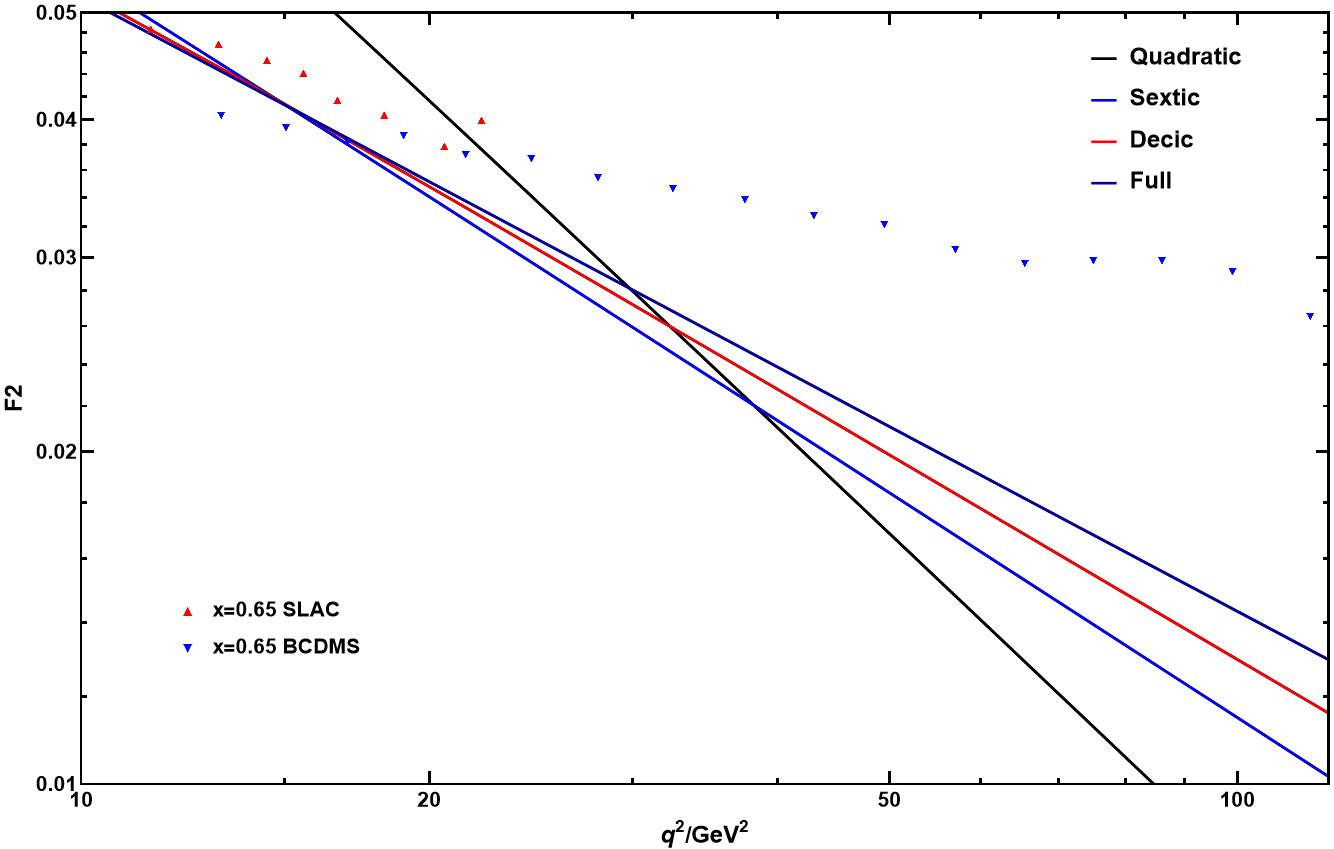}
\caption{Proton structure function $F_2(q^2,x)\times q^2$ obtained from the of the deformed AdS model for $x=0.65$. Black line - quadratic approximation (\ref{goodpotential1}), sextic approximation - blue line (\ref{sexticpotential}), for decic approximation - red line (\ref{decicpotential}), the full solution is given by the dark blue line (\ref{fullpotentials}).}
\label{fig7}
\end{figure}

    Here, in this section, we compare the behavior of the proton structure functions obtained in the previous section for different order of approximations of the chiral potentials. We use the values of the parameters presented in Tables [\ref{tab1},\ref{tab2}]. We plot in Figures \ref{fig5}, \ref{fig6} and \ref{fig7} our results for $x=0.85$, $x=0.75$ and $x=0.65$, for the quadratic, sextic, and decic approximations and the full potential in each graph. 

    The comparison of the results for the structure function $F_2(q^2,x)\times q^2$ for $x=0.85$, 0.75 and 0.65, for the different approximations and the full potentials presented in Figs. \ref{fig5}, \ref{fig6} and \ref{fig7}, clearly shows that the greater the value of $x$, the better are the results with respect with experimental data. This is  expected
    once we are using a supergravity prescription for  DIS \cite{Polchinski:2002jw}, which works better for larger values of $x$. 
    
    In particular, for $x=0.85$, plotted in Fig. \ref{fig5}, one sees that the sextic and decic approximations and the full chiral potentials give good results when compared with experimental data, while the quadratic approximation presents only reasonable results. 

    For $x=0.75$, the results for the sextic approximation is good for small values of $q^2$ up to 25 GeV$^2$, while the decid approximation and the full potentials are good up to $q^2\le 40\,{\rm GeV}^2$, and the quadratic approximation does not present a good fit to experimental data. 

    Finally, for $x=0.65$, the lower value of $x$ that we examine, the full potential and the sextic and decic approximations only give reasonable results when compared with experimental data for $q^2\le 20\,{\rm GeV}^2$, while the quadratic approximation presents analytical solutions, which in this case are very poor.

\section{Discussion and conclusions}\label{sec6}

    In this work we study the fermionic deep inelastic scattering in a deformed AdS model, considering  systematic approximations for the chiral potentials in order to preserve the isospectrality of the left and right fermionic modes, via SUSY quantum mechanics construction of a superpotential, as discussed in \cite{Costa-Silva:2026eqp}. 

    These approximations allow us to compare our results with ones coming from the soft wall model. At the quadratic approximation, with fermionic potentials given by (\ref{goodpotential1}), we obtained analytical solutions for the fermionic fields, and constructed the fermionic DIS to obtain the proton structure functions, that are also analytical in this model. We compared our results with the soft wall model and showed that with some transformations between the parameters of the models, and choosing $m_5 =1$, the models present identical results for DIS. The numerical results for the quadratic approximation can be compared to this in soft wall \cite{Braga:2011wa}, where the authors worked with $x\geq0.8$ only. As we have shown in Figs. \ref{fig5}, \ref{fig6}, and \ref{fig7}, the quadratic approximation (and consequently the soft wall model) fails for $x=0.75$ or lower. This behaviour is expected since the supergravity prescription used here \cite{Polchinski:2002jw} works better for higher values of $x$. 

    An important point of our work is that for high-order approximations of the chiral potentials, the results for the structure functions are better when compared with experimental data. This is closely related to the success of the DIS from deformed AdS model presented in \cite{FolcoCapossoli:2020pks}. 
    
     In summary,  our supergravity DIS prescription works well for large values of $x$. In order to holographically describe DIS for lower values, one needs to go further and consider the scattering of strings in our AdS/QCD model; we left this for future work.

\section*{Acknowledgements}
RACS is supported by Coordenação de Aperfeiçoamento de Pessoal de Nível Superior (CAPES). HBF is partially supported by Conselho Nacional de Desenvolvimento Cient\'{\i}fico e Tecnol\'{o}gico (CNPq) under grant  310346/2023-1, and Fundação Carlos Chagas Filho de Amparo à Pesquisa do Estado do Rio de Janeiro (FAPERJ) under grant E-26/204.095/2024.%


\appendix

\section{Detailed solution of the quadratic approximation}
\label{AppendA}

 The quadratic approximation to the chiral potentials, Eq. \eqref{goodpotential1} substituted in the Schrödinger-like equation \eqref{deformedpsieq}, gives
        \begin{equation}\label{deformedpsieq2}
        \bigg[-\frac{d^2}{dz^2}+\frac{m_5(m_5\mp1)}{z^2}+km_{5}\bigg(m_5\pm\frac{1}{2}\bigg)+\frac{m_5^2k^2}{4}z^2\bigg]\psi^{n}_{R/L}(z)=M_{n}^{2}\psi_{R/L}^{2}(z).
    \end{equation}

    Let us obtain the solutions for this equation and the corresponding spectra. We will find an appropriate trial solution examining its behavior in the limits of $z$. As $z$ approaches zero, the centrifugal term $m_5(m_5\mp1)/z^2$ dominates over the other terms, and the differential equation reduces to
    \begin{equation}
        \bigg(-\frac{d^2}{dz^2}+\frac{m_5(m_5\mp1)}{z^2}\bigg)\psi(z)\approx0.
    \end{equation}
    If we consider a power-law trial solution $\psi(z)\sim z^{\alpha}$, we obtain
    \begin{equation}
        \alpha(\alpha-1)=m_5(m_5\mp1).
    \end{equation}
    This dictates the power-law behavior near the origin. For very large values of $z$, the harmonic oscillator potential $m_5kz^2/2$ becomes the dominant term, so we have approximately
    \begin{equation}
        \bigg(-\frac{d^2}{dz^2}+\frac{(m_5kz)^2}{4}\bigg)\psi(z)\approx0, 
    \end{equation}
    which presents as solutions $e^{\pm (m_5kz)^2/4}$. To ensure normalizability, we select the decaying branch, i.e., $e^{-(m_5kz)^2/4}$. Combining these two exact asymptotic limits, we propose an \textit{ansatz} with an unknown function $F(u)$, with $u=m_5kz^2/2$, to be determined
    \begin{equation}
        \psi(z)=z^{\alpha}e^{-(m_5kz)^2/4}F(m_5kz^2/2).
    \end{equation}
    For simplicity, let us define some terms in the original equation: $C=m_5(m_5\mp1)$, $\omega=m_5k/2$, $\Tilde{M}_n^2=M_n^2-m_5k(m_5\pm1/2)$. With these definitions, the equation becomes
    \begin{equation}
        \bigg(-\frac{d^2}{dz^2}+\frac{C}{z^2}+\omega^2z^2\bigg)\psi(z)=\Tilde{M}_n^2\psi(z).
    \end{equation}

    Rewriting this in terms of $u=\omega z^2$, we obtain the confluent hypergeometric equation
    \begin{equation}
        uF''+\bigg(\alpha+\frac{1}{2}-u\bigg)F'-\lambda F=0,
    \end{equation}
    where $\lambda=\Tilde{M}_n^2/4\omega - \alpha/2 -1/4$. The solutions are the confluent hypergeometric functions,
    \begin{equation}
        F(u)=C_1\, _1F_1(\lambda,\alpha+1/2,u)+C_2\,\mathcal{U}(\lambda,\alpha+1/2,u).
    \end{equation}

    In order to obtain normalizable wave functions over the domain $z\in [0,\infty)$, it must not diverge as $z\rightarrow\infty$. In general, $_1F_1(a,b,u) \sim e^{u}$ as $u\rightarrow\infty$. To guarantee the normalizability of the wave function, the hypergeometric series must truncate into a polynomial, and it occurs when $\lambda$ is a non positive integer, i.e.,
    \begin{equation}
        \lambda=-n\,;\qquad(n=0,1,2,..).
    \end{equation}

    This relation implies
    \begin{equation}
        \frac{M_n^2}{2m_5k}-m_5-\frac{1}{2}=n,
    \end{equation}
    and we can obtain the spectrum 
    \begin{equation}\label{deformedspectra}
        M_n^2=2m_5k\bigg(n+m_5+\frac{1}{2}\bigg).
    \end{equation}

    Since $\lambda = -n$, there is a correspondence between the first Kummer hypergeometric function and the associated Laguerre polynomials,
    \begin{equation}
        _1F_1( -n,b,u)=\frac{n!}{(b)_n}L_n^{b-1}(u),
    \end{equation}
    where $(b)_n$ is the Pochhammer symbol. So, obtaining the square integrable function over this domain,
    \begin{equation}
        \psi_n(z)=N_nz^{\alpha}e^{-\omega z^2/2}L_n^{\alpha-1/2}(\omega z^2).
    \end{equation}

    Computing the normalization constant, and substituting the correspondent relations for each chiral mode, we obtain the solutions for the fermionic fields in the quadratic approximation of the deformed AdS model.
    
    \begin{equation}\label{psiR}
\psi_R^n(z)
=
\sqrt{
\frac{
\Gamma(n+1)
}{
\Gamma\left(n+m_5+\frac12\right)
}
}\sqrt{2\left(\dfrac{m_5 k}{2}\right)^{m_5+\frac12}}z^{m_5}
e^{-\frac{m_5 k}{4}z^2}L_n^{m_5-\frac12}\left(\frac{m_5 k}{2}z^2\right),
\end{equation}

\begin{equation}\label{psiL}
\psi_L^n(z)=
\sqrt{
\frac{\Gamma(n+1)
}{
\Gamma\left(n+m_5+\frac32\right)
}
}\sqrt{
2\left(\dfrac{m_5 k}{2}\right)^{m_5+\frac32}}z^{m_5+1}
e^{-\frac{m_5 k}{4}z^2}
L_n^{m_5+\frac12}
\left(\frac{m_5 k}{2}z^2\right).
\end{equation}


\bibliographystyle{elsarticle-num}
\bibliography{Bibliography}

\end{document}